\documentclass[fleqn,twoside,dvips]{mystyle}
\usepackage[tbtags]{amsmath}
\usepackage{amssymb,bm}
\usepackage{graphicx}

\usepackage{txfonts}

\usepackage{cite}

\def\<{\langle}
\def\>{\rangle}

\begin{document}

\title{A Cosmological basis for $E=mc^2$}

\author{
{\large\textbf{\textsf{Fulvio Melia}}}\\[1em]
{\rm Department of Physics, The Applied Math Program, and Department of Astronomy, \\
The University of Arizona, Tucson AZ 85721, USA}\\[1em]
{\rm E-mail: fmelia@email.arizona.edu}
}
\pacs{04.20.Jb, 95.30.Sf, 98.80.Es, 98.80.Jk}

\date{Submitted: 10 April 2018}

\iopabs{The Universe has a gravitational horizon with a radius
$R_{\rm h}=c/H$ coincident with that of the Hubble sphere. This
surface separates null geodesics approaching us from those receding,
and as free-falling observers within the Friedmann-Lema\^itre-Robertson-Walker
spacetime, we see it retreating at proper speed $c$, giving rise to
the eponymously named cosmological model $R_{\rm h}=ct$. As of today,
this cosmology has passed over 20 observational tests, often better
than $\Lambda$CDM. The gravitational radius $R_{\rm h}$ therefore
appears to be highly relevant to cosmological theory, and in this
paper we begin to explore its impact on fundamental physics.
We calculate the binding energy of a mass $m$ within the horizon
and demonstrate that it is equal to $mc^2$. This energy is
stored when the particle is at rest near the observer, transitioning
to a purely kinetic form equal to the particle's escape energy
when it approaches $R_{\rm h}$. In other words, a particle's
gravitational coupling to that portion of the Universe with which
it is causally connected appears to be the origin of rest-mass energy.
\\[5pt]
Keywords: General Relativity: exact solutions, Relativity and Gravitation, 
Observational Cosmology, Mathematical and Relativistic Aspects of Cosmology
}

\maketitle

\section{Introduction}
The Universe has a gravitational horizon with radius $R_{\rm h}=c/H$,
where $H$ is the Hubble constant, coincident with the better known Hubble
sphere \cite{Melia:2007,Melia:2016a,Melia:2017a,Melia:2017b,Melia:2009}. Unlike
its counterpart in the Schwarzschild and Kerr metrics, however, $R_{\rm h}$
is time-dependent so this surface may or may not eventually turn into
an event horizon in the asymptotic future depending on the cosmic fluid's
equation of state. The gravitational horizon was formally introduced in
ref.~\cite{Melia:2007}, though an unidentified predecessor appeared almost
a century ago in de Sitter's \cite{deSitter:1917} own account
of his now famous solution. In the intervening years, the choice of
coordinates for which $R_{\rm h}$ appears explicitly in the metric
was lost following the popularization of the comoving frame,
principally by Friedmann \cite{Friedmann:1923}. In this paper, we shall have
occasion to use the Friedmann-Lema\^itre-Robertson-Walker
(FLRW) metric written in terms of both sets of coordinates.

The role played by $R_{\rm h}$ in any interpretation of the data is so
important that a cosmological model based on its properties, known
as the $R_{\rm h}=ct$ universe
\cite{Melia:2003,Melia:2007,Melia:2016a,Melia:2017a,Melia:2009,Melia:2012a},
has already passed over 20 observational tests, typically better than
$\Lambda$CDM. A summary of the model comparisons may be found in
Table 1 of ref.~\cite{Melia:2017c}. An example of the impact $R_{\rm h}$
can have on our understanding of cosmological features is the role it
played in resolving the question concerning whether or not cosmological
redshift represents a new kind of time dilation, separate from the more
conventional gravitational and Doppler effects. The answer
is no---cosmological redshift is simply the product of these two
\cite{Melia:2012b}, better known as the `lapse' function in other
applications of general relativity.

The concept of a gravitational radius in cosmology is not always
easy to grasp because the observational evidence suggests the Universe
is infinite. We are embedded within it, however, and the gravitational
influence between us and another spacetime point depends solely on the
intervening energy content. This may be understood quite easily in the
context of the Birkhoff theorem \cite{Birkhoff:1923} and its corollary
(see also refs.~\cite{Weinberg:1972,Melia:2007}). As such, every observer
or particle---no matter where they are---is surrounded by a gravitational
horizon a proper distance $R_{\rm h}$ away because the rest of the
Universe exterior to this surface has a vanishing gravitational
influence on the interior.

Such a limitation to our causal connectedness suggests a possible
impact on fundamental physics. In this paper, we begin to
examine this issue by asking a very basic---yet profound---question
concerning the nature of rest-mass energy---specifically, whether it
may be related in some way to a particle's binding energy within the
gravitational horizon.

We should emphasize at the outset that we are here making a clear
distinction between the origin of inertia, i.e., rest mass, $m$, and the
nature of rest-mass energy, $mc^2$. As far as we know today, the Higgs
mechanism, with its SU(2) internal symmetry group, endows inertia to
elementary particles that couple to the Higgs field
\cite{EnglertBrout:1964,Higgs:1964}. Why inertia is associated
with an energy $mc^2$ is a different question.

\section{The Friedmann-Lema\^itre-Robertson-Walker Metric}
We begin with the FLRW metric for a
spatially homogeneous and isotropic three-dimensional space, scaled
by the expansion factor $a(t)$:
\begin{equation}
ds^2=c^2\,dt^2-a^2(t)\left[{dr^2\over 1-kr^2}+
r^2(d\theta^2+\sin^2\theta\,d\phi^2)\right]\;.
\end{equation}
\vskip 0.1in\noindent
The comoving coordinates used in this expession include the cosmic
time $t$, an appropriately scaled radial coordinate $r$, and angular
coordinates $\theta$ and $\phi$. The geometric factor $k$ is $+1$
for a closed universe, $0$ for a flat universe, and $-1$ for an open
universe. The high-precision measurements available today
\cite{Planck:2014a,Planck:2014b} suggest that the Universe is flat,
so we will assume the value $k=0$ throughout this paper.

As we proceed through this discussion, we shall see that
$(ct,r,\theta,\phi)$ are the coordinates of a {\it free-falling}
observer, analogous to a counterpart in the Schwarzschild or Kerr
spacetimes. But for the latter, it has also been very useful to recast
the metric in a form relevant to an {\it accelerated} observer---one
who is at rest with respect to the central mass---and we shall similarly
follow this procedure in the cosmological context. To do this,
we introduce the proper radius, $R(t)\equiv a(t)r$, often used to express
the distance that changes along with the expansion of the Universe. This
proper distance $R(t)$ is a direct consequence of Weyl's postulate
applied to an isotropic universe \cite{Weyl:1923}, i.e., that no
two worldlines in a cosmology satisfying the Cosmological principle
should ever cross following the big bang---other than from local peculiar
motions---which requires every distance in FLRW to be expressible as the
product of an unchanging comoving length $r$ and a universal,
position-independent function of time, $a(t)$.

We shall follow the procedure introduced in refs.~\cite{Melia:2009,Melia:2017b}
to rewrite the FLRW metric in terms of $R(t)$. Writing the expansion factor in the form
\begin{equation} 
a(t)=e^{f(t)}\;,
\end{equation}
we put
\begin{equation}
r=Re^{-f}\;,
\end{equation}
so that
\begin{equation}
dr=e^{-f}\,dR-\dot{f}r\,dt\;.
\end{equation}
The metric in Equation~(1) thereby becomes
\begin{equation}
ds^2 = c^2\,dt^2\left[1-\left({R\dot{f}\over c}\right)^2\right]+
2\left({R\dot{f}\over c}\right)c\,dt\,dR -dR^2-R^2\,d\Omega^2\;,
\end{equation}
where, for convenience, we have defined
\begin{equation}
d\Omega^2\equiv d\theta^2+\sin^2\theta\,d\phi^2\;.
\end{equation}
Now introducing the function
\begin{equation}
\Phi\equiv 1-\left(\frac{R}{R_{\rm h}} \right)^2\;,
\end{equation}
which will signal the dependence of the metric coefficients $g_{tt}$ and $g_{RR}$ on
the proximity of $R(t)$ to the gravitational radius $R_{\rm h}$, the first
two terms in Equation~(5) may be rewritten as follows:
\begin{eqnarray}
c^2\,dt^2-a^2\,dr^2&\hskip-0.1in=\hskip-0.1in&\Phi\left[c^2\,dt^2-\Phi^{-1}\,dR^2+2c\,dt\,
\left({R\dot{f}\over c}\right)\Phi^{-1}\,dR\right]\nonumber \\
&\hskip-0.1in=\hskip-0.1in&\Phi\left[c\,dt+\left({R\dot{f}\over c}\right)\Phi^{-1}\,dR\right]^2-\Phi^{-1}\,dR^2\;.
\end{eqnarray}
We now consider the line element along the worldlines of particular observers,
those that have $t$ as their proper time from one location to the next, i.e.,
comoving observers, as it turns out. Introducing the proper speed $\dot{R}\equiv dR/dt$
along these worldlines, we may then complete the square in Equation~(8) and
write Equation~(5) as
\begin{equation}
ds^2= \Phi\left[1 + \left(\frac{R}{R_{\rm h}} \right)\Phi^{-1}
{\dot{R}\over c}  \right]^2c^2\,dt^2 - \Phi^{-1}{dR^2}-R^2\,d\Omega^2
\end{equation}
Some may see a similarity between this form of the metric and
that used to derive the Oppenheimer-Volkoff equations for the interior of
a star \cite{Oppenheimer:1939,Misner:1964} except, of course, that the
latter is static, whereas both $R(t)$ and $R_{\rm h}(t)$ are functions
of $t$ in FLRW.

\section{Binding Energy}
Let us now define the 4-momentum of a particle
\begin{equation}
p^\mu \equiv (E/c,p^R,p^\theta,p^\phi)\;,
\end{equation}
where $E$ is its energy, and $p^i$ are the usual spatial components,
and consider the invariant contraction $p^\mu p_\mu$. For the metric
coefficients in Equation~(9), one has
\begin{equation}
\Phi \left[1+\left({R\over R_{\rm h}}\right)\Phi^{-1}{\dot{R}\over c}\right]^2
\left({E\over c}\right)^2-\Phi^{-1}\left(m\dot{R}\right)^2=K^2\;,
\end{equation}
where $K$ is a constant (i.e., a scalar) yet to be determined, and we
have assumed purely radial motion with $p^\theta=p^\phi=0$ and
\begin{equation}
p^R=m\dot{R}\;,
\end{equation}
in terms of the particle's rest mass $m$. Note that no additional factor,
such as a time dilation, appears in Equation~12 because the cosmic time $t$,
used in the derivative, is also the local proper time at every spacetime point
in the cosmic fluid. In the Appendix, we demonstrate that the contraction
of $p^\mu$ with itself, based on the definitions in Equations~(10) and (12), is
a scalar and a constant in the spacetime described by Equation~(9).
Equation~(11) thus expresses the particle's energy $E$ in terms of its
momentum $m\dot{R}$ everywhere in the medium, starting from the observer's
location at the origin ($R=0$) all the way to the gravitational horizon at
$R_{\rm h}$.

Let us re-write it in a somewhat more conventional form,
\begin{equation}
E^2={(cK)^2\Phi+(mc)^2{\dot{R}}^2\over
\left[\Phi+\left({R\over R_{\rm h}}\right){\dot{R}\over c}\right]^2}\;,
\end{equation}
and first consider what happens at the horizon. There $R=R_{\rm h}$
and $\dot{R}=c$, while $\Phi=0$. Clearly,
\begin{equation}
E(R_{\rm h}) = mc^2\;.
\end{equation}
But notice that this value comes---not from $K$, which one would naively
have assumed ab initio---but rather from the momentum transitioning to
its relativistic limit, i.e., $p^R\rightarrow mc$, while the contribution
from $K$ itself gets redshifted away completely as a result of
$\Phi\rightarrow 0$ when $R\rightarrow R_{\rm h}$. This result is quite
remarkable because it tells us that the particle's escape energy as it
nears the gravitational horizon is what we would normally call its
rest-mass energy $mc^2$. The emphasis here is on the phrase `escape
energy' because this value of $E$ is entirely due to $p^R$ at $R_{\rm h}$.

Assuming that the particle has no peculiar velocity at $R<R_{\rm h}$,
we may also write
\begin{equation}
m\dot{R}=mc\left({R\over R_{\rm h}}\right)\;,
\end{equation}
and therefore the general expression for the total energy is
\begin{equation}
E^2=(mc^2)^2\left[1-\left({R\over R_{\rm h}}\right)^2\right]\left({K\over mc}\right)^2
+(mc^2)^2\left({R\over R_{\rm h}}\right)^2\;.
\end{equation}
A quick inspection of Equation~(9) shows that in the $R_{\rm h}=ct$
universe, the metric coefficients $g_{tt}$ and $g_{RR}$ are
time-independent. This is because both $R(t)$ and $R_{\rm h}$
are proportional to $t$. And as is well known in general relativity
\cite{Weinberg:1972}, energy is conserved along a particle geodesic---here
represented by Equation~(15)---when the spacetime metric is independent
of time \cite{Killing:1892}. In addition, the fact that the
$R_{\rm h}=ct$ universe has zero active mass, i.e., $\rho+3p=0$
\cite{Melia:2016a,Melia:2017a}, means that the particle
experiences zero net acceleration, so it cannot gain or
lose energy from the background, and therefore $E$ in Equation~(16)
must be constant within the framework of $R_{\rm h}=ct$. But according
to this energy conservation equation, $E$ can be constant only for one
particular value of $K$, and that is $K=mc$, in which case
\begin{equation}
E=mc^2
\end{equation}
everywhere and at all times.

This equally remarkable result tells us that the total energy $E$ can remain
constant even though $p^R$ increases from $0$ at the origin to its maximum
value $mc$ at $R_{\rm h}$. We interpret this to mean that
the particle's binding energy $mc^2$ at the origin is gradually converted
into kinetic energy as its proper distance from us nears our gravitational
horizon, and $E$ becomes entirely kinetic when $R=R_{\rm h}$, but always
equal to $mc^2$.

Notice also that we began our comparison of the gravitational horizon in
cosmology with its counterpart in Schwarzschild and Kerr by emphasizing
the fact that $R_{\rm h}$ changes with time. Yet none of the results,
particularly Equations~(14) and (17), are affected by this. Even as
$R_{\rm h}$ increases with time, $E$ always remains constant and
$p^R$ depends only on the {\it ratio} $R/R_{\rm h}$. So the value
$E=mc^2$ and its transition from binding to kinetic energy (via
Eqs.~13 and 15) remain valid forever. As long as a proton's mass
has remained constant in time, its rest-mass energy today is identical
to its rest-mass energy minutes after the big bang.

\section{Discussion}
The quantity $E=mc^2$ may be interpreted as a gravitational binding
energy because, according to the observer at the origin, this is how much
energy the particle would need to free itself from its gravitational coupling
to the Universe within $R_{\rm h}$.
The region exterior to $R_{\rm h}$ does not participate in this gravitational
interaction. It is apparently this $E$ that is gradually converted into
kinetic energy (in the form of $p^R$), reaching its ``escape" value
$p^Rc=(mc)c$ at the gravitational radius $R_{\rm h}$. Note that in this
sense, $mc^2$ is literally the binding energy required to climb out of
the gravitational potential well.

Mathematical consistency with these ideas is ensured by the invariance
of the contracted 4-momentum vector, $p^\mu p_\mu$, which tells us exactly
how the energy is changing in terms of the particle's momentum. The physical
descriptions we provide here inform our understanding of what is happening,
but ultimately it is the invariance of the scalar $K$ that yields the dependence
of $p^R$ on $R$. We do not actually have to calculate $E$ from the gravitational
interaction itself. This is already done for us through the presence of $\Phi(R)$
in the metric. In other words, the redshift effect represented by $\Phi$
accounts for the gravitationl attraction the particle feels to the rest of
the Universe within $R_{\rm h}$.

A more subtle point has to do with why the particle's inertial mass is
proportional (or even equal) to its gravitational mass. We do not attempt
to broach this subject here, but as is well known, this is the basis for the
Principle of Equivalence in general relativity. With it, we may use the particle's
inertia to characterize the strength of its gravitational interaction with the
surrounding medium, so it is legitimate for us to ask what its gravitational
binding energy is in terms of $m$. Of course, this is the reason we can
interpret $mc^2$ as a gravitational binding energy in the first place.
If inertia were unrelated to the gravitational mass, then there would be
no physical reason at all for us to argue that the rest energy associated
with $m$ should have anything to do with gravity.

When discussing such concepts, it clearly matters who the observer
is. From the perspective of an observer fixed at the origin of the
coordinates $(ct,R,\theta,\phi)$, the Universe is not static. Every
particle moves away from him at the Hubble speed, $\dot{R}$, which
increases steadily and reaches $c$ when $R=R_{\rm h}$. From his
perspective, the cosmic fluid has a total energy commensurate with
its momentum $p^R$. Thus, if the origin of a particle's rest energy
$mc^2$ were independent of its recessional velocity, the Hubble flow
would be progressively more energetic as $R\rightarrow R_{\rm h}$
which, as we have seen, is not confirmed by the invariance of
$p^\mu p_\mu$. So for this particular observer, the quantity $mc^2$
represents a blend of stored and kinetic energy, which transitions
to $p^Rc$ completely at the gravitational horizon.

When viewed in the comoving frame, however, the cosmic fluid
is always at rest (other than for peculiar velocities that do not
contribute to the true Hubble flow). Observers in this frame
therefore see only the energy $E=mc^2$ corresponding to $p^r=0$.
Strictly speaking, there is a different free-falling frame at
each new location, so the particle's rest energy is measured by
different observers at different spacetime points. It is this
switching from one observer to the next that replaces the
variation of $p^R$ with distance in the accelerated frame.

Finally, it may be worth mentioning that the approach we have followed
here in deriving our result has some overlap with the method commonly 
used to infer the mass-energy of so-called cosmological black holes. 
Unlike static black holes in a flat spacetime background, real black 
holes must necessarily be embedded within an expanding FLRW metric (see, e.g.,
refs~\cite{Nolan:1998,Nolan:1999,Kaloper:2010,Firouzjaee:2018,Moradi:2017,Lake:2011}).
Modeling these extended bodies in a curved background introduces
various degrees of coupling between their mass-energy and the geometry
of the Universe at large, notably its apparent (or gravitational)
horizon \cite{Melia:2018}. This in turn affects their dynamics and
their own horizon. While this topic does not directly refer to the
nature of rest-mass energy per se, the relationship between the enclosed
energy of cosmological black holes and the type of background metric arises
from the same gravitational interaction within a causally connected 
region that we have invoked to calculate the binding energy
of a fundamental particle within the Universe's horizon. Some issues
revolving around how to best define masses and energy for cosmological
black holes still remain unresolved, but the steps taken to couple the 
Kerr (or Schwarzschild) and FLRW metrics are based on similar physical
principles that we have used in this paper.

\section{Conclusion}
The identification of rest-mass energy with the binding energy inside
our gravitational horizon is thus quite compelling. Indeed, our argument
is based entirely on core principles in general relativity. Were rest-mass
energy due to something else, one would need to explain---within the framework
of this theory---why the total energy at $R_{\rm h}$ is not greater than
$mc^2$, in spite of the fact that $p^R\rightarrow mc$.

The success of the $R_{\rm h}=ct$ cosmology in providing such an elegant,
accessible explanation for the origin of rest-mass energy adds to its
credentials as a viable description of nature. Its principal divergence
from $\Lambda$CDM is that it does not have a horizon problem, so it does
not have or need inflation to account for the uniformity of the microwave
background across the sky \cite{Melia:2013}. Without inflation
\cite{Guth:1981,Linde:1982}, the standard model could not survive, yet even
after four decades of study, we still do not have a complete, self-consistent
understanding of the inflaton field (see, e.g., refs.~\cite{Ijjas:2013,Ijjas:2014}).
Perhaps this too is an indication that inflation never happend, pointing
to the $R_{\rm h}=ct$ universe as the only viable cosmology. Additional
high-precision tests are underway \cite{Melia:2016b}, and we may
have a definitive answer within a matter of years.

\section{Appendix}
In this appendix, we demonstrate that the contraction $p^\mu p_\mu$,
with the four-momentum defined in Equations~(10) and (12), is a
scalar and a constant in the spacetime given by the metric in Equation~(9).
In doing so, we recall the discussion concerning the selected worldlines
with proper speed $\dot{R}\equiv dR/dt$ preceding this equation, and
we simplify the procedure by invoking the condition of zero peculiar
motion everywhere, i.e., $\dot{r}=0$. As such, $\dot{R}=\dot{a}r=HR$,
where $H$ is the Hubble constant $H\equiv \dot{a}/a$. In addition,
it is trivial to see that $\dot{R_{\rm h}}=c$ in the $R_{\rm h}=ct$
universe, since $R_{\rm h}\equiv c/H$ (see, e.g., refs.~\cite{Melia:2009,Melia:2012a}).
Since the Universe is isotropic and homogeneous, the geodesics are
radial so, with zero peculiar velocities, we may also write the four-velocity
$U^\mu\equiv dX^\mu/d\tau=dX^\mu/dt$, where $X^\mu=(ct,R,\theta,\phi)$, as
\begin{equation}
U^\mu=(c,\dot{R},0,0)\;.
\end{equation}

Let us now consider the time evolution of $p^\mu p_\mu$, with $p^\mu\equiv mU^\mu$.
That is, we shall proceed to evaluate the derivative
\begin{equation}
{d\over dt}\left(p^\mu p_\mu\right)=m^2{dU^\mu\over dt}U_\mu+m^2U^\mu{dU_\mu\over dt}\;.
\end{equation}
With the four-velocity in Equation~(18), and its covariant analogue
\begin{equation}
U_\mu\equiv g_{\mu\nu}U^\nu\;,
\end{equation}
in which only the metric coefficients
\begin{equation}
g_{tt}\equiv \Phi\left[1 + \left(\frac{R}{R_{\rm h}} \right)\Phi^{-1}
{\dot{R}\over c}  \right]^2
\end{equation}
and
\begin{equation}
g_{RR}\equiv - \Phi^{-1}
\end{equation}
are non-zero, Equation~(19) becomes
\begin{eqnarray}
\hskip-0.3in {d\over dt}\left(p^\mu p_\mu\right)&=&2m^2\dot{R}\left(-\Phi^{-1}\right)\ddot{R}-
m^2{\dot{R}}^2{d\over dt}\Phi^{-1}+\nonumber \\
&\null&\hskip-0.4in m^2c^2{d\over dt}\left\{\Phi\left[1+\left({R\over R_{\rm h}}\right)\Phi^{-1}{\dot{R}\over c}\right]^2\right\}\;.
\end{eqnarray}
In evaluating the right-hand side of this equation, it will be
helpful to see that
\begin{equation}
{d\over dt}\Phi={d\over dt}\Phi^{-1}=0\;,
\end{equation}
and that
\begin{equation}
{d\over dt}\left[1+\left({R\over R_{\rm h}}\right)\Phi^{-1}{\dot{R}\over c}\right]^2=
2\Phi^{-2}{R\over R_{\rm h}}{\ddot{R}\over c}\;.
\end{equation}

Therefore,
\begin{eqnarray}
\hskip-0.25in {d\over dt}\left(p^\mu p_\mu\right)&=&-m^2\Phi^{-1}{d\over dt}{\dot{R}}^2+
2m^2c^2\Phi^{-1}{R\over R_{\rm h}}{\ddot{R}\over c}\nonumber \\
&=&-m^2\Phi^{-1}{d\over dt}{\dot{R}}^2+m^2\Phi^{-1}{d\over dt}{\dot{R}}^2\nonumber \\
&=& 0\;,
\end{eqnarray}
so the contraction of the four-momentum $p^\mu$ is clearly a scalar and
a constant in this spacetime.

\ack

I am grateful to the Instituto de Astrof\'isica de Canarias in Tenerife and
to Purple Mountain Observatory in Nanjing, China for their hospitality while
part of this work was carried out. I also acknowledge partial support from the
Chinese Academy of Sciences Visiting Professorships for Senior International
Scientists under grant 2012T1J0011, and from the Chinese State Administration
of Foreign Experts Affairs under grant GDJ20120491013.

\end{document}